\PassOptionsToPackage{table,xcdraw}{xcolor}
\documentclass[11pt,letterpaper]{article}
\usepackage[margin=1in]{geometry}
\usepackage{amsmath,epsfig}
\usepackage{stfloats}
\usepackage{breqn}
\usepackage{xcolor}
\usepackage{float}
\usepackage{url}
\usepackage{subcaption}
\usepackage{amsmath,amssymb,amsfonts}
\usepackage{graphicx}
\usepackage{booktabs}
\usepackage{caption}
\usepackage[style=ieee]{biblatex}

\definecolor{lightgray}{gray}{0.9}

\def\x{{\mathbf x}}
\def\X{{\mathbf X}}
\def\Z{{\mathbf Z}}
\def\m{{\mathbf m}}
\def\M{{\mathbf M}}
\def\Rmat{{\mathbf R}}
\def\D{{\mathcal D}}
\def\Q{{\mathbf Q}}
\def\Y{{\mathbf Y}}

\title{SpectraLift: Physics-Guided Spectral-Inversion Network for Self-Supervised Hyperspectral Image Super-Resolution}

\author{
  Ritik Shah\\
  University of Massachusetts\\
  Amherst, MA 01003\\
  \texttt{rgshah@umass.edu}
  \and
  Marco F. Duarte\\
  University of Massachusetts\\
  Amherst, MA 01003\\
  \texttt{mduarte@umass.edu}
}

\begin{document}

\maketitle

\begin{abstract}
High–spatial-resolution Hyperspectral Images (HSI) are essential for applications such as remote sensing and medical imaging, yet HSI sensors inherently trade spatial detail for spectral richness. Fusing High–spatial-Resolution Multispectral Images (HR-MSI) with Low–spatial-Resolution Hyperspectral Images (LR-HSI) is a promising route to recover fine spatial structures without sacrificing spectral fidelity. Most state of the art methods for HSI-MSI fusion demand Point Spread Function (PSF) calibration or ground truth High-spatial-Resolution HSI (HR-HSI), both of which are impractical to obtain in real world settings. We present SpectraLift, a fully self-supervised framework that fuses LR-HSI and HR-MSI inputs using only the MSI’s Spectral Response Function (SRF). SpectraLift trains a lightweight per-pixel Multi-Layer Perceptron (MLP) network using ($i$) a synthetic Low–spatial-Resolution Multispectral Image (LR-MSI) obtained by applying the SRF to the LR-HSI as input, ($ii$) the LR-HSI as the target, and ($iii$) an $\ell_1$ spectral reconstruction loss between the estimated and true LR-HSI as the optimization objective. At inference, SpectraLift uses the trained network to map the HR-MSI pixel-wise into a HR-HSI estimate. SpectraLift converges in minutes, is agnostic to spatial blur and resolution, and outperforms state-of-the-art methods on PSNR, SAM, SSIM, and RMSE benchmarks.
\end{abstract}

\section{Introduction}
\label{sec:intro}
Hyperspectral images (HSI) capture hundreds of narrow spectral bands, enabling precise material discrimination. However, dispersive optics and photon splitting impose a trade-off: richer spectral resolution entails coarser spatial detail, which blurs small features, exacerbates mixed-pixel effects, and degrades classification accuracy.
Fusion-based Hyperspectral Super-Resolution (HSI-SR) addresses this trade-off by combining a LR-HSI with a HR-MSI of the same scene. Early approaches (pan-sharpening, Bayesian inference, matrix factorization, tensor decomposition) leverage handcrafted priors but often struggle with complex spectral structures. Recent deep learning-based methods deliver high reconstruction fidelity but depend on scarce ground-truth HR-HSI, intricate architectures, and opaque “black-box” mappings that hinder transparency.

Unsupervised and hybrid schemes incorporate physics priors (e.g., PSF models, sparsity, low-rank constraints) to reduce supervision and improve interpretability. Yet they typically demand precise PSF knowledge—impractical given variable sensor blur. State-of-the-art unsupervised methods try to estimate the PSF instead of strictly requiring precise knowledge of it. However, blurring in hyperspectral sensors is caused due to unknown and variable optical characteristics such as lens diffraction and sensor-specific aberrations, resulting in an inherent ambiguity. It is thus extremely difficult to estimate the PSF that caused the blur, which limits real-world deployment and raises concerns about model robustness and trustworthiness.

We propose SpectraLift, a self-supervised, lightweight framework that fuses LR-HSI and HR-MSI using only the multispectral sensor's SRF. Note that the SRF is manufacturer-defined, well-documented, and routinely used for radiometric and atmospheric correction; common approximations such as Gaussian estimations also work well in practice. A synthetic LR-MSI is generated for training purposes by applying the SRF to the LR-HSI. This LR-MSI and the original LR-HSI serve (pixelwise) as the input-output pair to train a compact MLP that we call the Spectral Inversion Network (SIN). The SIN learns an implicit MSI to HSI spectral mapping for the training images by minimizing an $\ell_{1}$ loss between the estimated and true LR-HSI. At inference, the trained SIN applies its learned spectral mapping to output a HR-HSI from a HR-MSI input.

By formulating HSI-SR as per-pixel spectral inversion, SpectraLift can avoid PSF estimation and any HR-HSI supervision  while being agnostic to spatial blur and resolution. SpectraLift training converges quickly and yields fully interpretable spectral mappings. Through extensive experiments on multiple benchmark datasets, we show that SpectraLift consistently outperforms state-of-the-art supervised and unsupervised methods in PSNR, SAM, SSIM, and RMSE, while remaining lightweight.

During the preparation of this manuscript we became aware of~\cite{SSSR}, which also employs an MLP for spectral mapping that is optimized with an MSE loss. However, their training strategy generates LR-MSI by spatially blurring and downsampling HR-MSI -- thus requiring precise PSF knowledge. In contrast, SpectraLift relies solely on the SRF and trains with a pixel-wise $\ell_{1}$ loss, yielding a more robust formulation and superior results than~\cite{SSSR}, as shown in Section~\ref{ssec:experiments}. 
 
\vspace{-2mm}
\section{Proposed Method}
\label{sec:method}
\vspace{-1mm}

\begin{figure*}[!ht]
  \centering
  \begin{subfigure}[b]{\textwidth}
    \centering
    \includegraphics[width=\textwidth]{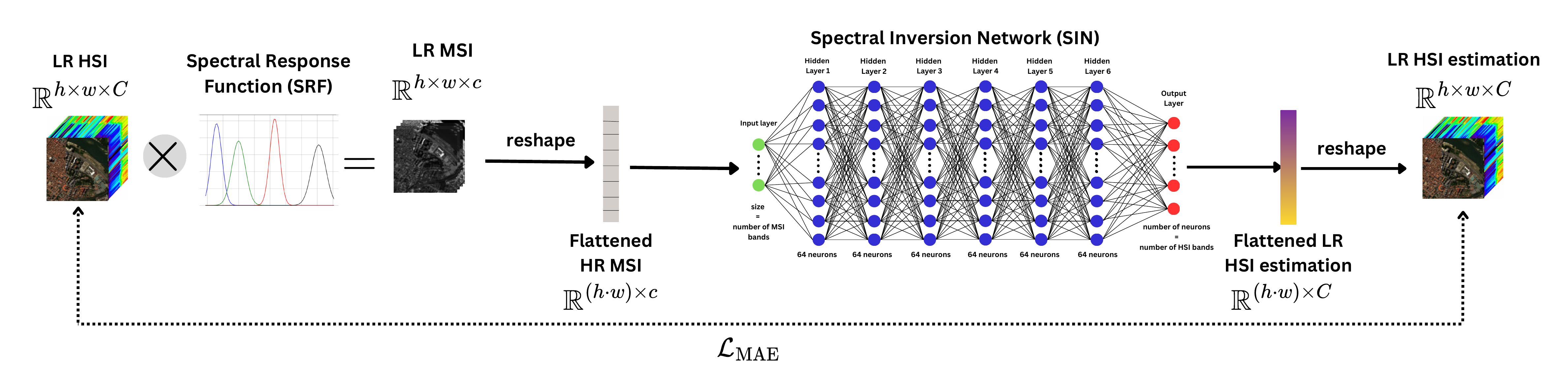}
    \label{fig:spectralift_train}
  \end{subfigure}
  \vspace{-2mm}
  \begin{subfigure}[b]{0.8\textwidth}
    \centering
    \includegraphics[width=\textwidth]{SpectraLift_inference.pdf}
    \label{fig:spectralift_infer}
  \end{subfigure}
  \vspace{-2mm}
  \caption{The SpectraLift pipelines. {\em Top:} self-supervised training of the Spectral Inversion Network (SIN) via SRF-based spectral inversion. {\em Bottom:} pixel-wise inference on HR‐MSI with the trained SIN to produce the super resolved High-spatial-Resolution Hyperspectral Image (HR HSI)}.
  \label{fig:spectralift}
\end{figure*}

The MSI sensor's SRF acts as a band compression operator, mapping high-dimensional HSI spectra to lower-dimensional MSI measurements. SpectraLift, illustrated in Figure~\ref{fig:spectralift}, formulates HSI-SR as the self-supervised, per-pixel inversion of this spectral compression. Given only a pair of LR-HSI and HR-MSI, as well as the SRF (or its approximation), a compact MLP learns to reverse the SRF-induced degradation and reconstruct the HSI from an MSI input. SpectraLift uses only the LR-HSI image for training and the HR-MSI image for inference.

Let
$\Y  \in \mathbb{R}^{h \times w \times C}$
be the observed LR-HSI with \(C\) bands, and
$\M \in \mathbb{R}^{H \times W \times c}$ be
the HR-MSI with \(c < C\) bands. Under classical sensor models, these relate to the unknown HR-HSI
$\X \in \mathbb{R}^{H \times W \times C}$
via spatial and spectral degradations:
$\Y \approx \D_r\bigl(\mathcal{H}(\X, \Q)\bigr)$ and $\M \approx \X \times \Rmat$,
where \(\mathcal{H}(\X,\Q)\) denotes spatial convolution of the HSI image \(\X\) with the PSF \(\Q\), \(\D_r(\X)\) denotes spatial downsampling of the HSI \(\X\) by a factor \(r\), and \(\X \times \Rmat\) denotes the product of the HSI \(\X\) with the MSI SRF \(\Rmat\in\mathbb{R}^{C\times c}\) along the spectral (third) mode, i.e., for $i=1,\ldots,H$, $j = 1,\ldots,W$, $m=1,\ldots,c$, we have
\begin{align*}
    \M_{ijm}
    =
    \sum_{n=1}^{C} \X_{ijn}\,\Rmat_{nm}.
\end{align*}

\subsection{Spectral Inversion Network (SIN)}

Denote a given input MSI pixel from the image \(\M\) by \({\mathbf m} = \M_{ij:}\) and the corresponding output HSI pixel estimate by \(\hat{\mathbf x}\). SIN impements a per-pixel mapping \(\hat{\x}=f_\Theta(\m)\), where \(\Theta\) denotes the set of SIN parameters, from an MSI spectrum \(\m\in\mathbb{R}^c\) to an HSI spectrum \(\hat{\x}\in\mathbb{R}^C\). The layers in SIN can be mathematically described as: 
  \begin{align*}
    \mathbf{x}^{(0)} &= \m,\\
    \mathbf{x}^{(1)} &= \phi_1\bigl(\mathbf{x}^{(0)}\bigr), \\
    \mathbf{x}^{(2)} &= \phi_2\bigl(\mathbf{x}^{(1)}\bigr) + \mathbf{x}^{(1)}, \\
    \mathbf{x}^{(3)} &= \phi_3\bigl(\mathbf{x}^{(2)}\bigr), \\
    \mathbf{x}^{(4)} &= \phi_4\bigl(\mathbf{x}^{(3)}\bigr) + \mathbf{x}^{(2)}, \\
    \mathbf{x}^{(5)} &= \phi_5\bigl(\mathbf{x}^{(4)}\bigr), \\
    \mathbf{x}^{(6)} &= \phi_6\bigl(\mathbf{x}^{(5)}\bigr) + \mathbf{x}^{(4)}, \\
    \hat{\mathbf{x}} &= g_\theta\bigl(\mathbf{x}^{(6)}\bigr),
\end{align*}
  where each \(\phi_i\) is a fully connected layer with 64 neurons and leaky ReLU activation, and \(g_\theta\) is a fully connected output layer with linear activation. Residual/skip connections are incorporated in every other layer to enhance model expressiveness without increasing architectural complexity. Their inclusion leads to consistently improved spectral reconstruction and smoother convergence.

\noindent\textbf{Training and Implementation:} We synthesize the LR-MSI \(\Z=\Y\times\Rmat\) and optimize
\[ \min_{\Theta}\;\frac{1}{hw}\sum_{i=1}^h\sum_{j=1}^w\bigl\|f_\Theta(\Z_{ij:}) - \Y_{ij:}\bigr\|_1, \]
training the SIN to invert the spectral degradation under an \(\ell_1\) loss. 
SpectraLift is implemented using the TensorFlow framework and trained with the Adam optimizer. For learning rate scheduling, we use the One-Cycle Learning Rate policy for the Washington DC Mall and Kennedy Space Center datasets, as it accelerates convergence and promotes stability. For the Pavia University, Pavia Center, Botswana, and University of Houston datasets, we employ cosine annealing with restarts, which yielded superior convergence in these cases. 

We tuned the scheduler parameters -- specifically, the initial, maximum, and final learning rates for the One-Cycle policy, and the maximum and minimum learning rates for cosine annealing with restarts. The learning rates used in our experiments are provided in our pre-executed Jupyter notebooks. These notebooks, available in our GitHub repository (see section \ref{ssec:experiments}), contain the exact configurations used for each experiment.

\noindent\textbf{Inference:} At test time, the trained SIN is applied to each HR-MSI pixel \(\M_{ij:}\), yielding the pixels of the estimated HR-HSI \(\hat{\X}\).

\subsection{Rationale for Per-Pixel Formulation}
SpectraLift’s design deliberately treats each pixel independently during spectral inversion. This choice is motivated by several practical considerations. 

\noindent \textbf{HSI-MSI temporal misalignment:} Spatial priors typically assume co-registered HSI and MSI data, which is often unavailable in real-world scenarios due to temporal misalignments or platform differences, e.g., UAV vs.\ satellite acquisitions. Thus, moving objects may shift between acquisitions. This breaks the pixel-by-pixel correspondence assumed by many state of the art fusion methods and causes visible artifacts in the super-resolved outputs. SpectraLift entirely avoids this issue: it trains on synthetic LR-MSI derived solely from the LR-HSI and, at inference, processes each HR-MSI pixel independently, without ever requiring co-registration. Thus, no artifacts appear due to temporal offsets.

\noindent \textbf{Dependence on PSF:} Spatially coupled models are highly sensitive to unknown PSF variations, which can lead to artifacts when applied across diverse sensors or acquisition geometries. This design simplifies the training process and dramatically reduces computational overhead without sacrificing spectral fidelity, as evidenced by our results in Section \ref{ssec:experiments}. 

\noindent \textbf{Spatial Resolution and Blur Agnosticism:}
Because training uses only individual pixels, and both the input and target during training have very similar blurring, \(f_\Theta\) has no mechanism to learn or depend on spatial structure. Consequently, the same learned mapping applies unchanged to any spatial sampling grid and optical acquisition setup, endowing SpectraLift with genuine spatial-resolution and quality agnosticism.

\subsection{SIN Insights and Limitations}
\textbf{Exact SRF inversion:}
Under ideal conditions (\(c \ge C\)), \(\Rmat\) admits a true inverse \(\Rmat^\dagger\) (Moore-Penrose pseudo-inverse). In practice \(c < C\), making inversion in \(\mathbb{R}^C\) ill-posed; however, real-world spectra lie on a low-dimensional non-linear manifold of intrinsic dimension \(s \ll c,C\). The MLP learns a stable approximation \(f_\Theta \approx \Rmat^\dagger\) restricted to this manifold.

\noindent \textbf{Physics-Guided Interpretability:}
Our sole physics assumption is the SRF \(\Rmat\), which is routinely provided by MSI manufacturers. This contrasts sharply with PSF-dependent unsupervised methods (and in particular~\cite{SSSR}), since accurate PSF estimation is difficult or impossible in practice. With only a lightweight network and a clear \(\ell_1\) loss, SpectraLift converges rapidly and yields directly interpretable spectral inversion filters, avoiding the opaque ``black-box'' nature of most current state-of-the-art models.

\noindent \textbf{Limitations:} A limitation of SpectraLift arises when the HR-MSI input contains only a single spectral band (e.g., a monochrome image). In this extreme case, predicting a full hyperspectral signature from a scalar quantity is extremely ill-posed. In contrast, other state-of-the-art methods can exploit spatial context or impose assumptions about the similarity of spectra in neighboring pixels of the high-resolution output. While such assumptions can improve performance for highly degraded inputs, they may also reduce generalization when the input retains more spectral information. Furthermore, we note that single band HR images are rarely acquired in real world image fusion scenarios. Since SpectraLift only learns to invert the scene dependent SRF, it must be retrained for each new LR-HSI/HR-MSI pair. The SIN trained on one scene cannot be directly transferred to another, limiting the potential for cross-scene generalization which is a common limitation among state-of-the-art unsupervised/self-supervised HSI-SR methods. 

\section{Experiments and Analysis}
\label{ssec:experiments}
We evaluate SpectraLift on synthetic and real-world benchmarks. We compare against eight state-of-the-art baselines: four unsupervised (SSSR~\cite{SSSR}, MIAE~\cite{MIAE}, C2FF~\cite{C2FF}, SDP~\cite{SDP}) and four supervised (GuidedNet~\cite{GuidedNet}, FeINFN~\cite{FeINFN}, FusFormer~\cite{FusFormer}, MIMO-SST~\cite{MIMO}).   
All datasets, precomputed results, and detailed instructions for replication are available in our GitHub repository, which can be accessed at \url{https://github.com/ritikgshah/SpectraLift}. We have made every effort to ensure reproducibility by providing pre-executed Jupyter notebooks, Python scripts for end-to-end runs, configuration files, and environment replication scripts for seamless setup. This enables reviewers and practitioners to verify results and explore extensions with minimal friction.

\subsection{Quality Metrics} For synthetic data, we evaluate SpectraLift using six widely adopted metrics that capture complementary aspects of reconstruction quality in hyperspectral image super-resolution:

\begin{itemize}
    \item {\bf Root Mean Squared Error (RMSE):} Measures the average pixel-wise difference between the reconstructed hyperspectral image \(\hat{\X}\) and the ground truth \(\X\). Lower RMSE indicates higher reconstruction fidelity.
    \[
    \mathrm{RMSE}(\X,\hat{\X}) = \sqrt{\frac{1}{HWC}\sum_{i=1}^{H}\sum_{j=1}^{W}\sum_{k=1}^{C} \bigl(\hat{\X}_{ijk} - \X_{ijk}\bigr)^2}.
    \]

    \item {\bf Peak Signal-to-Noise Ratio (PSNR):} Quantifies the ratio between the maximum possible pixel value and the power of the reconstruction error, expressed in decibels (dB). Higher PSNR indicates better perceptual quality.
    \[
    \mathrm{PSNR}(\X,\hat{\X}) = 10 \log_{10}\left(\frac{\mathrm{MAX}^2}{\mathrm{RMSE}(\X,\hat{\X})^2}\right),
    \]
    where \(\mathrm{MAX}\) is the maximum possible pixel value (e.g., \(2^{16}-1\) for 16-bit images).

    \item {\bf Structural Similarity Index Measure (SSIM):} Assesses perceptual similarity by comparing luminance, contrast, and structure between \(\hat{\X}\) and \(\X\). Values close to 1 indicate high structural similarity.
    \[
    \mathrm{SSIM}(\X, \hat{\X}) = \frac{(2\mu_X\mu_{\hat{X}} + c_1)(2\sigma_{X\hat{X}} + c_2)}{(\mu_X^2 + \mu_{\hat{X}}^2 + c_1)(\sigma_X^2 + \sigma_{\hat{X}}^2 + c_2)},
    \]
    where \(\mu\), \(\sigma^2\), and \(\sigma_{X\hat{X}}\) are means, variances, and covariances, and \(c_1, c_2\) are small constants to stabilize the denominator.

    \item {\bf Universal Image Quality Index (UIQI):} Measures similarity in terms of luminance, contrast, and structure. Values range from \(-1\) to 1, with higher values indicating better quality.
    \[
    \mathrm{UIQI}(\X,\hat{\X}) = \frac{4\sigma_{X\hat{X}}\mu_X\mu_{\hat{X}}}{\left(\sigma_X^2 + \sigma_{\hat{X}}^2\right)\left(\mu_X^2 + \mu_{\hat{X}}^2\right)}.
    \]

    \item {\bf Erreur Relative Globale Adimensionnelle de Synth\`{e}se (ERGAS):} Provides a global indication of the relative error, normalized by the mean reflectance, and is commonly used in remote sensing. Lower ERGAS indicates higher reconstruction accuracy.
    \[
    \mathrm{ERGAS}(\X,\hat{\X}) = 100~\frac{r}{s}\sqrt{\frac{1}{C}\sum_{k=1}^{C}\frac{\mathrm{RMSE}_k(\X, \hat{\X})^2}{\mu_k^2}},
    \]
    where \(r/s\) is the ratio of spatial resolutions between \(\hat{\X}\) and \(\X\), and \(\mathrm{RMSE}_k(\X, \hat{\X})\) and \(\mu_k\) are the RMSE and mean of band \(k\), respectively.

    \item {\bf Spectral Angle Mapper (SAM):} Computes the mean spectral angle (in degrees) between estimated and ground-truth spectral vectors at each pixel. It measures spectral similarity, with smaller angles indicating better fidelity. To ensure numerical stability, the arccos argument is clipped to avoid numerical issues.

    \[
    \mathrm{SAM}(\X, \hat{\X}) = \frac{1}{HW} \sum_{i=1}^{H}\sum_{j=1}^{W} \left( \frac{180}{\pi} \arccos\left( \min\left( \frac{\langle \X_{ij:}, \hat{\X}_{ij:} \rangle}{\|\X_{ij:}\|_2 \|\hat{\X}_{ij:}\|_2 + \epsilon}, 1 - \delta \right) \right) \right),
    \]
    
    where $\X_{ij:}$ and $\hat{\X}_{ij:}$ are the spectral vectors at pixel $(i,j)$, and $\epsilon$ and $\delta$ are small constants to avoid division by zero and numerical overflow, respectively (e.g., $\epsilon=10^{-8}$, $\delta=10^{-9}$).

\end{itemize}

We also profile execution time, number of parameters, inference peak GPU memory used and inference FLOPs to assess model complexity. For the inference GPU memory used and inference FLOPs we consider these for a single forward pass of the inputs through the model. In the tables below, we specify whether the best value of a metric is lower ($\downarrow$) or higher ($\uparrow$).

\subsection{Synthetic Datasets}
\begin{figure*}[!ht]
    \centering
    \includegraphics[width=\textwidth]{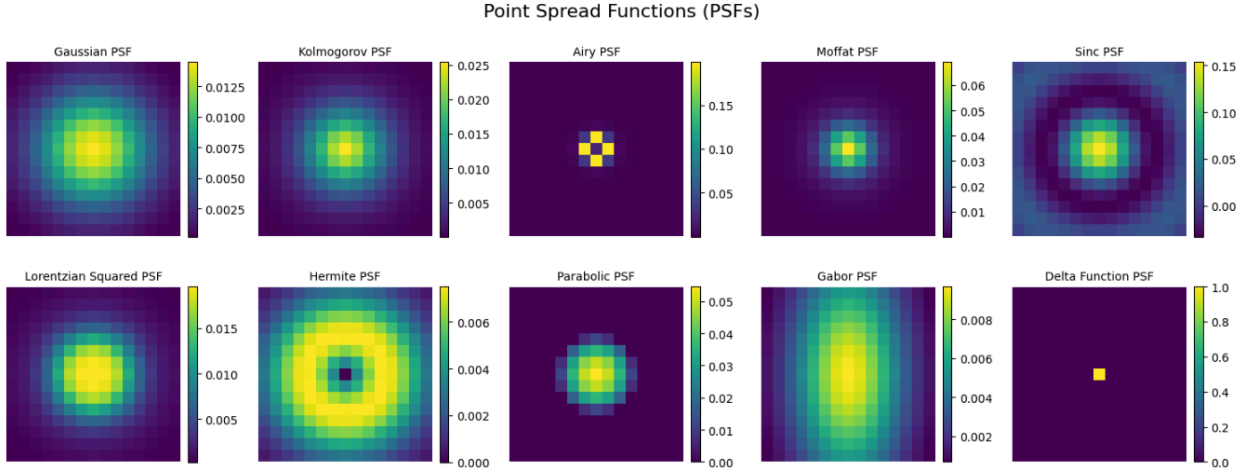} 
    \caption{Point Spread Functions used for Synthetic LR HSI generation}
    \label{fig:psf}
\end{figure*}

Following Wald's protocol~\cite{wald}, for each ground-truth HSI (GT): Washington DC Mall (DC), Kennedy Space Center (KSC), Botswana, Pavia University (Pavia U), and Pavia Center (Pavia), we generate LR-HSI by first convolving the GT with one of ten PSFs shown in Figure \ref{fig:psf} (Gaussian, Kolmogorov, Airy, Moffat, Sinc, Lorentzian Squared, Hermite, Parabolic, Gabor, Delta), all with kernel size of (15,15). We then downsample the convolution output by \(r\in\{4,8,16,32\}\), and finally add Gaussian white noise with SNR matched to \(r\): (4, 35 dB), (8, 30 dB), (16, 25 dB), (32, 20 dB), yielding \(10\times4=40\) LR-HSIs for each GT. HR-MSI are synthesized by applying SRFs for \(c\in\{1,3,4\}\) (IKONOS), \(c=8\) (WorldView-2), and \(c=16\) (WorldView-3), followed by adding Gaussian white noise with SNR=40 dB, producing 5 HR-MSIs for each GT. 

We consider 80 LR-HSI/HR-MSI pairs per GT: 10 PSFs $\times$ 8 representative $(r,c)$ configurations \{(4,4), (8,4), (16,4), (32,4), (8,1), (8,3), (8,8), (8,16)\}. We apply all the degradations to the GT normalized between [0,1].
Supervised methods are trained on 75\% crops and tested on 25\% crops for each LR-HSI-HR-MSI input pair. Spatial dimensions for the training and testing crops are constrained to be  multiples of 32, ensuring compatibility with models that require integer downsampling without residual pixels. Unsupervised methods use full images but report metrics on the same 25\% test regions. To support a more fair comparison across supervised and unsupervised approaches, the latter are also given access to the PSF and SRF used for LR-HSI and HR-MSI generation.

Tables \ref{tab:dc_quality}-\ref{tab:botswana_quality} report mean quality and complexity metrics over these 80 LR-HSI/HR-MSI pairs for the Washington DC Mall, Kennedy Space Center, Pavia University, Pavia Center, and Botswana benchmarks, respectively, while Tables \ref{tab:dc_quality_without_1_band}-\ref{tab:botswana_quality_without_1_band} show the same quality metrics when the single-band HR image is omitted. We do not include the model complexity metrics in tables \ref{tab:dc_quality_without_1_band}-\ref{tab:botswana_quality_without_1_band} since they have almost no difference compared to the ones reported in tables \ref{tab:dc_quality}-\ref{tab:botswana_quality} . Across all five synthetic datasets, SpectraLift outperforms unsupervised baselines on almost all quality measures. Supervised methods enjoyed a modest advantage -- they were trained on a 75\% crop of the same scene (distinct from the 25\% test region), allowing them to learn scene-specific spectral behaviors. Despite lacking this advantage, on naturally occurring scenes (Washington DC Mall, Kennedy Space Center, Botswana) SpectraLift even surpasses most supervised approaches, attaining the highest PSNR and lowest SAM on DC, state-of-the-art RMSE, PSNR, SSIM, ERGAS  and SAM on KSC, and the highest SSIM and lowest SAM on Botswana. Only MIMO-SST edges ahead, but with a more complex method relying on "black box" models with knowledge of the scene specific spectra. On urban scenes with more man-made materials (Pavia University, Pavia Center), scene-specific spectral information gives supervised models like MIMO-SST and FeINFN an edge, but SpectraLift still outperforms unsupervised baselines, demonstrating its advantage when comparing methods with equal training knowledge. When we omit the single-band HR image -- reflecting more realistic operational scenarios -- SpectraLift rises to the top on the natural scenes: it secures best RMSE, PSNR, SSIM and SAM on DC (Table~\ref{tab:dc_quality_without_1_band}), best results for all metrics (except UIQI) on KSC (Table~\ref{tab:ksc_quality_without_1_band}), and best SSIM, UIQI, and SAM on Botswana (Table~\ref{tab:botswana_quality_without_1_band}), while remaining highly competitive on the urban benchmarks. These results underscore its effectiveness when the HR input has more than a single band.

\begin{table*}[!htbp]
  \centering
  \setlength\tabcolsep{4pt}          
  \renewcommand{\arraystretch}{1.1}  
  \captionsetup{font=small, justification=centering}
  \caption[Quality measures]{Quality measures for the Washington DC Mall dataset: Mean value of 80 LR-HSI/HR-MSI configurations (\textbf{best in bold}, \textit{second best in italics}; supervised methods are shaded)}, 
  \label{tab:dc_quality}
  
  \resizebox{\columnwidth}{!}{%
  \begin{tabular}{|l|c|c|c|c|c|c|c|c|c|c|}
    \hline
    \textbf{Method} & \textbf{RMSE $\downarrow$} & \textbf{PSNR $\uparrow$} & \textbf{SSIM $\uparrow$} 
    & \textbf{UIQI $\uparrow$} & \textbf{ERGAS $\downarrow$} & \textbf{SAM $\downarrow$} 
    & \textbf{Time (s) $\downarrow$} & \textbf{Params (M) $\downarrow$} & \textbf{FLOPs (G) $\downarrow$} 
    & \textbf{GPU Mem (MB) $\downarrow$} \\
    \hline
    MIAE & 0.03611 & 30.28 & 0.934 & 0.968 & 5.33  & 5.21  & 212.69  & \textbf{0.0218} & 8.46    & 707.54 \\
    \hline
    C2FF & 0.02681 & 34.74 & 0.960 & \textit{0.975} & \textit{4.73}  & 3.53  & \textit{78.54} & 0.0979 & \textit{7.29}    & 948.88 \\
    \hline
    SDP & 0.03029 & 31.99 & 0.939 & 0.907 & 16.59 & 4.29  & 799.75  & 6.4546 & 511.52  & 5370.37 \\
    \hline
    SSSR & 0.05024 & 27.98 & 0.908 & 0.938 & 7.49  & 5.83  & \textbf{67.24}   & \textit{0.0334} & \textbf{0.000067} & \textit{286.31} \\
    \hline
    \rowcolor{lightgray}
    GuidedNet & 0.03494 & 29.98 & 0.919 & 0.872 & 30.30 & 4.81  & 398.07  & 6.7111 & 178.41  & \textbf{81.52} \\
    \hline
    \rowcolor{lightgray}
    FeINFN & 0.02558 & 32.66 & 0.966 & 0.956 & 7.83  & 3.79  & 3018.80 & 3.7021 & 251.70  & 982.47 \\
    \hline
    \rowcolor{lightgray}
    FusFormer & 0.02555 & 32.55 & 0.876 & 0.688 & 17.34 & 3.89  & 11110.07 & 0.1883 & 946.28  & 6328.36 \\
    \hline
    \rowcolor{lightgray}
    MIMO-SST & \textbf{0.02193} & \textit{35.21} & \textbf{0.969} & \textbf{0.982} & \textbf{2.92}  & \textit{3.29}  & 376.34  & 2.1879 & 56.37   & 393.74 \\
    \hline
    SpectraLift & \textit{0.02344} & \textbf{35.96} & \textit{0.967} & 0.969 & 5.87  & \textbf{3.22}  & 91.20   & 0.0336 & 26.28   & \textit{286.31} \\
    \hline
  \end{tabular}}
\end{table*}

\begin{table*}[!htbp]
  \centering
  \footnotesize
  \setlength\tabcolsep{4pt}
  \renewcommand{\arraystretch}{1.1}
  \captionsetup{font=small, justification=centering}
  \caption{Quality measures for the Kennedy Space Center dataset: Mean value of 80 LR HSI-HR MSI configurations (\textbf{best in bold}, \textit{second best in italics}; supervised methods are shaded).}
  \label{tab:ksc_quality}
  \resizebox{\columnwidth}{!}{%
  \begin{tabular}{|l|c|c|c|c|c|c|c|c|c|c|}
    \hline
    \textbf{Method}
      & \textbf{RMSE $\downarrow$}
      & \textbf{PSNR $\uparrow$}
      & \textbf{SSIM $\uparrow$}
      & \textbf{UIQI $\uparrow$}
      & \textbf{ERGAS $\downarrow$}
      & \textbf{SAM $\downarrow$}
      & \textbf{Time (s) $\downarrow$}
      & \textbf{Params (M) $\downarrow$}
      & \textbf{FLOPs (G) $\downarrow$}
      & \textbf{GPU Mem (MB) $\downarrow$} \\
    \hline
    MIAE & 0.04855 & 26.41 & 0.907 & \textbf{0.962} & 8.54  & \textit{8.88}  & 189.24  & 0.0996  & 31.12   & 711.06 \\
    \hline
    C2FF & \textit{0.04540} & \textit{26.97} & \textit{0.922} & 0.956 & \textit{8.50}  & 8.90  & \textit{77.24}   & 0.0913  & \textit{5.42}    & 707.95 \\
    \hline
    SDP & 0.04659 & 26.73 & 0.906 & 0.949 & 8.63  & 9.65  & 643.65  & 6.1322  & 388.55  & 4209.39 \\
    \hline
    SSSR & 0.05767 & 25.05 & 0.851 & 0.930 & 9.54  & 10.08 & \textbf{70.70}   & \textbf{0.0284}  & \textbf{0.000057} & \textit{211.06} \\
    \hline
    \rowcolor{lightgray}
    GuidedNet & 0.04951 & 26.11 & 0.913 & 0.956 & 9.17  & 9.25  & 374.44  & 6.0301  & 176.43  & \textbf{71.32} \\
    \hline
    \rowcolor{lightgray}
    FeINFN & 0.05013 & 26.02 & 0.902 & 0.932 & 9.07  & 10.77 & 2206.34 & 3.6520  & 261.73  & 984.15 \\
    \hline
    \rowcolor{lightgray}
    FusFormer & 0.05124 & 25.82 & 0.888 & 0.921 & 9.26  & 11.57 & 6976.85 & 0.1811  & 787.98  & 6322.09 \\
    \hline
    \rowcolor{lightgray}
    MIMO-SST & 0.04660 & 26.68 & 0.917 & 0.941 & 8.70  & 9.47  & 305.21  & 2.1361  & 56.79   & 363.68 \\
    \hline
    SpectraLift & \textbf{0.04472} & \textbf{27.10} & \textbf{0.925} & \textit{0.957} & \textbf{8.41}  & \textbf{8.70}  & 92.24   & \textit{0.0327}  & 20.42   & \textit{211.06} \\
    \hline
  \end{tabular}}
\end{table*}

\begin{table*}[!htbp]
  \centering
  \footnotesize
  \setlength\tabcolsep{4pt}
  \renewcommand{\arraystretch}{1.1}
  \captionsetup{font=small, justification=centering}
  \caption{Quality measures for the Pavia University dataset: Mean value of 80 LR HSI-HR MSI configurations (\textbf{best in bold}, \textit{second best in italics}; supervised methods are shaded).}
  \label{tab:pavia_u_quality}
  \resizebox{\columnwidth}{!}{%
  \begin{tabular}{|l|c|c|c|c|c|c|c|c|c|c|}
    \hline
    \textbf{Method}
      & \textbf{RMSE $\downarrow$}
      & \textbf{PSNR $\uparrow$}
      & \textbf{SSIM $\uparrow$}
      & \textbf{UIQI $\uparrow$}
      & \textbf{ERGAS $\downarrow$}
      & \textbf{SAM $\downarrow$}
      & \textbf{Time (s) $\downarrow$}
      & \textbf{Params (M) $\downarrow$}
      & \textbf{FLOPs (G) $\downarrow$}
      & \textbf{GPU Mem (MB) $\downarrow$} \\
    \hline
    MIAE & 0.03775 & 29.62 & 0.906 & 0.984 & 3.64  & 5.01  & 99.99   & 0.0880  & 18.11   & 445.91 \\
    \hline
    C2FF & 0.03618 & 31.70 & 0.914 & 0.976 & 3.36  & 4.63  & \textit{73.24}  & 0.0590  & \textit{2.25}    & 293.47 \\
    \hline
    SDP & 0.03088 & 31.88 & 0.917 & 0.982 & 3.14  & 4.29  & 385.58  & 4.6211  & 192.66  & 2491.03 \\
    \hline
    SSSR & 0.05851 & 26.31 & 0.829 & 0.950 & 5.46  & 5.68  & \textbf{66.48}   & \textbf{0.0099}  & \textbf{0.000020} & \textit{81.49} \\
    \hline
    \rowcolor{lightgray}
    GuidedNet & 0.03060 & 30.80 & 0.920 & \textit{0.986} & 3.36  & 4.19  & 243.72  & 3.4673  & 68.62   & \textbf{22.63} \\
    \hline
    \rowcolor{lightgray}
    FeINFN & \textit{0.02789} & 31.82 & \textit{0.933} & 0.982 & 3.04  & \textit{4.18}  & 1187.35 & 3.4082  & 127.63  & 381.94 \\
    \hline
    \rowcolor{lightgray}
    FusFormer & 0.03017 & 31.00 & 0.929 & 0.983 & 3.32  & 4.22  & 5556.52 & 0.1460  & 471.06  & 6231.11 \\
    \hline
    \rowcolor{lightgray}
    MIMO-SST & \textbf{0.02236} & \textbf{34.34} & \textbf{0.945} & \textbf{0.989} & \textbf{2.49}  & \textbf{3.33}  & 221.63  & 1.8836  & 22.93   & 153.75 \\
    \hline
    SpectraLift & 0.03073 & \textit{32.79} & 0.928 & 0.980 & \textit{2.99}  & \textit{4.18}  & 91.03   & \textit{0.0279}  & 11.52   & \textit{81.49} \\
    \hline
  \end{tabular}}
\end{table*}

\begin{table*}[!htbp]
  \centering
  \footnotesize
  \setlength\tabcolsep{4pt}
  \renewcommand{\arraystretch}{1.1}
  \captionsetup{font=small, justification=centering}
  \caption{Quality measures for the Pavia Center dataset: Mean value of 80 LR HSI-HR MSI configurations (\textbf{best in bold}, \textit{second best in italics}; supervised methods are shaded).}
  \label{tab:pavia_c_quality}
  \resizebox{\columnwidth}{!}{%
  \begin{tabular}{|l|c|c|c|c|c|c|c|c|c|c|}
    \hline
    \textbf{Method}
      & \textbf{RMSE $\downarrow$}
      & \textbf{PSNR $\uparrow$}
      & \textbf{SSIM $\uparrow$}
      & \textbf{UIQI $\uparrow$}
      & \textbf{ERGAS $\downarrow$}
      & \textbf{SAM $\downarrow$}
      & \textbf{Time (s) $\downarrow$}
      & \textbf{Params (M) $\downarrow$}
      & \textbf{FLOPs (G) $\downarrow$}
      & \textbf{GPU Mem (MB) $\downarrow$} \\
    \hline
    MIAE & 0.04344 & 28.67 & 0.901 & 0.984 & 3.58  & 6.99  & 314.86  & 0.0878  & 68.30    & 1677.58 \\
    \hline
    C2FF & 0.02861 & 33.38 & 0.956 & 0.988 & 2.46  & 4.90  & \textit{85.77} & 0.0586  & \textit{8.43}     & 1094.10 \\
    \hline
    SDP & 0.02979 & 32.11 & 0.952 & 0.988 & 2.61  & 5.58  & 1085.04 & 4.6011  & 724.77   & 9375.14 \\
    \hline
    SSSR & 0.06399 & 25.67 & 0.839 & 0.958 & 5.26  & 6.28  & \textbf{67.55}   & \textbf{0.0098}  & \textbf{0.000020} & \textit{304.91} \\
    \hline
    \rowcolor{lightgray}
    GuidedNet & 0.02929 & 31.10 & 0.954 & \textit{0.992} & 2.68  & 4.98  & 723.75  & 3.4408  & 263.18   & \textbf{92.44} \\
    \hline
    \rowcolor{lightgray}
    FeINFN & 0.02532 & 32.48 & \textit{0.964} & \textit{0.992} & 2.33  & 4.83  & 5131.20 & 3.4048  & 490.85   & 1459.95 \\
    \hline
    \rowcolor{lightgray}
    FusFormer & \textit{0.02262} & 33.58 & \textbf{0.968} & \textbf{0.993} & \textit{2.09}  & \textit{4.49}  & 19432.89 & 0.1455  & 1884.16  & 6321.71 \\
    \hline
    \rowcolor{lightgray}
    MIMO-SST & \textbf{0.02168} & \textbf{35.02} & 0.963 & \textbf{0.993} & \textbf{2.01}  & \textbf{4.19}  & 706.33  & 1.8802  & 87.91    & 575.90 \\
    \hline
    SpectraLift & 0.02716 & \textit{33.88} & 0.958 & 0.988 & 2.36  & 4.80  & 90.84   & \textit{0.0278}  & 43.41    & \textit{304.91} \\
    \hline
  \end{tabular}}
\end{table*}

\begin{table*}[!htbp]
  \centering
  \footnotesize
  \setlength\tabcolsep{4pt}
  \renewcommand{\arraystretch}{1.1}
  \captionsetup{font=small, justification=centering}
  \caption{Quality measures for the Botswana dataset: Mean value of 80 LR HSI-HR MSI configurations (\textbf{best in bold}, \textit{second best in italics}; supervised methods are shaded).}
  \label{tab:botswana_quality}
  \resizebox{\columnwidth}{!}{%
  \begin{tabular}{|l|c|c|c|c|c|c|c|c|c|c|}
    \hline
    \textbf{Method}
      & \textbf{RMSE $\downarrow$}
      & \textbf{PSNR $\uparrow$}
      & \textbf{SSIM $\uparrow$}
      & \textbf{UIQI $\uparrow$}
      & \textbf{ERGAS $\downarrow$}
      & \textbf{SAM $\downarrow$}
      & \textbf{Time (s) $\downarrow$}
      & \textbf{Params (M) $\downarrow$}
      & \textbf{FLOPs (G) $\downarrow$}
      & \textbf{GPU Mem (MB) $\downarrow$} \\
    \hline
    MIAE & 0.01838 & 35.06 & 0.954 & \textit{0.997} & 1.75  & 1.77  & 162.78  & \textbf{0.0190}  & 7.09     & 552.00 \\
    \hline
    C2FF & 0.01572 & 36.51 & 0.970 & \textbf{0.998} & \textit{1.48}  & 1.59  & \textbf{74.82}   & 0.0776  & \textit{5.48}     & 716.48 \\
    \hline
    SDP & 0.01863 & 34.82 & 0.951 & 0.995 & 2.29  & 1.97  & 698.01  & 5.4788  & 416.77   & 4835.41 \\
    \hline
    SSSR & 0.02050 & 34.20 & 0.953 & 0.996 & 1.93  & 1.69  & \textit{86.12} & \textit{0.0194} & \textbf{0.000039} & \textit{208.97} \\
    \hline
    \rowcolor{lightgray}
    GuidedNet & 0.01970 & 34.25 & 0.927 & 0.986 & 4.11  & 1.93  & 439.14  & 4.7894  & 162.62   & \textbf{68.65} \\
    \hline
    \rowcolor{lightgray}
    FeINFN & 0.02250 & 33.44 & 0.946 & 0.995 & 1.95  & 2.53  & 2718.46 & 3.5485  & 266.91   & 919.15 \\
    \hline
    \rowcolor{lightgray}
    FusFormer & 0.02937 & 31.10 & 0.780 & 0.815 & 11.23 & 3.35  & 8333.27 & 0.1662  & 944.11   & 6299.68 \\
    \hline
    \rowcolor{lightgray}
    MIMO-SST & \textbf{0.01346} & \textbf{37.76} & \textit{0.973} & \textbf{0.998} & \textbf{1.35}  & \textit{1.54}  & 365.49  & 2.0289  & 53.87    & 367.98 \\
    \hline
    SpectraLift & \textit{0.01452} & \textit{37.04} & \textbf{0.974} & \textit{0.997} & 1.60  & \textbf{1.46}  & 93.97   & 0.0306  & 23.03    & 209.00\ \\
    \hline
  \end{tabular}}
\end{table*}

\begin{table}[t]
  \centering
  \small
  \renewcommand{\arraystretch}{1.1}
  \captionsetup{font=small, justification=centering}
  \caption{Quality measures for the Washington DC dataset without 1 band HR image: Mean value of 70 LR HSI-HR MSI configurations (\textbf{best in bold}, \textit{second best in italics}; supervised methods are shaded).}
  \label{tab:dc_quality_without_1_band}
  \begin{tabular}{|l|c|c|c|c|c|c|}
    \hline
    \textbf{Method}
      & \textbf{RMSE $\downarrow$}
      & \textbf{PSNR $\uparrow$}
      & \textbf{SSIM $\uparrow$}
      & \textbf{UIQI $\uparrow$}
      & \textbf{ERGAS $\downarrow$}
      & \textbf{SAM $\downarrow$} \\
    \hline
    MIAE & 0.03239 & 31.14 & 0.948 & 0.971 & 5.03  & 4.92 \\
    \hline
    C2FF & 0.01569 & 36.89 & \textit{0.988} & \textbf{0.989} & \textit{3.74}  & \textit{2.17} \\
    \hline
    SDP & 0.02214 & 33.49 & 0.961 & 0.914 & 16.29 & 3.49 \\
    \hline
    SSSR & 0.04269 & 29.16 & 0.926 & 0.947 & 7.01  & 4.75 \\
    \hline
    \rowcolor{lightgray}
    Guided Net & 0.03173 & 30.72 & 0.928 & 0.870 & 31.35 & 4.43 \\
    \hline
    \rowcolor{lightgray}
    FeINFN & 0.02142 & 33.72 & 0.978 & 0.959 & 7.58  & 3.42 \\
    \hline
    \rowcolor{lightgray}
    Fus Former & 0.02193 & 33.51 & 0.886 & 0.689 & 17.30 & 3.56 \\
    \hline
    \rowcolor{lightgray}
    MIMO-SST & \textit{0.01457} & \textit{37.00} & \textbf{0.991} & \textbf{0.989} & \textbf{2.16}  & 2.32 \\
    \hline
    Spectra Lift & \textbf{0.01266} & \textbf{38.22} & \textbf{0.991} & \textit{0.980} & 5.13  & \textbf{1.90} \\
    \hline
  \end{tabular}
\end{table}

\begin{table}[t]
  \centering
  \small
  \setlength\tabcolsep{2pt}
  \renewcommand{\arraystretch}{1.1}
  \captionsetup{font=small, justification=centering}
  \caption{Quality measures for the Kennedy Space Center dataset without 1 band HR image: Mean value of 70 LR HSI-HR MSI configurations (\textbf{best in bold}, \textit{second best in italics}; supervised methods are shaded).}
  \label{tab:ksc_quality_without_1_band}
  \begin{tabular}{|l|c|c|c|c|c|c|}
    \hline
    \textbf{Method}
      & \textbf{RMSE $\downarrow$}
      & \textbf{PSNR $\uparrow$}
      & \textbf{SSIM $\uparrow$}
      & \textbf{UIQI $\uparrow$}
      & \textbf{ERGAS $\downarrow$}
      & \textbf{SAM $\downarrow$} \\
    \hline
    MIAE        & 0.04736 & 26.62 & 0.915 & \textbf{0.963} & 8.40  & 8.75 \\
    \hline
    C2FF        & \textit{0.04247} & \textit{27.45} & \textit{0.936} & 0.959 & \textit{8.14}  & \textit{8.38} \\
    \hline
    SDP         & 0.04386 & 27.16 & 0.923 & 0.953 & 8.31  & 9.16 \\
    \hline
    SSSR        & 0.05608 & 25.31 & 0.857 & 0.933 & 9.29  & 9.55 \\
    \hline
    \rowcolor{lightgray}
    Guided Net  & 0.04851 & 26.28 & 0.920 & 0.957 & 9.06  & 9.09 \\
    \hline
    \rowcolor{lightgray}
    FeINFN      & 0.04926 & 26.16 & 0.908 & 0.932 & 8.98  & 10.74 \\
    \hline
    \rowcolor{lightgray}
    Fus Former  & 0.05059 & 25.93 & 0.892 & 0.921 & 9.19  & 11.54 \\
    \hline
    \rowcolor{lightgray}
    MIMO-SST    & 0.04450 & 27.03 & 0.933 & 0.943 & 8.46  & 9.09 \\
    \hline
    Spectra Lift & \textbf{0.04178} & \textbf{27.58} & \textbf{0.939} & \textit{0.960} & \textbf{8.05}  & \textbf{8.15} \\
    \hline
  \end{tabular}
\end{table}

\begin{table}[t]
  \centering
  \small
  \setlength\tabcolsep{2pt}
  \renewcommand{\arraystretch}{1.1}
  \captionsetup{font=small, justification=centering}
  \caption{Quality measures for the Pavia University dataset without 1 band HR image: Mean value of 70 LR HSI-HR MSI configurations (\textbf{best in bold}, \textit{second best in italics}; supervised methods are shaded).}
  \label{tab:pavia_u_quality_without_1_band}
  \begin{tabular}{|l|c|c|c|c|c|c|}
    \hline
    \textbf{Method}
      & \textbf{RMSE $\downarrow$}
      & \textbf{PSNR $\uparrow$}
      & \textbf{SSIM $\uparrow$}
      & \textbf{UIQI $\uparrow$}
      & \textbf{ERGAS $\downarrow$}
      & \textbf{SAM $\downarrow$} \\
    \hline
    MIAE        & 0.03348 & 30.50 & 0.924 & 0.986 & 3.29  & 4.81 \\
    \hline
    C2FF        & 0.02485 & 33.55 & 0.947 & 0.987 & 2.63  & 3.55 \\
    \hline
    SDP         & 0.02170 & 33.47 & 0.950 & 0.990 & 2.54  & 3.49 \\
    \hline
    SSSR        & 0.05040 & 27.38 & 0.855 & 0.959 & 4.99  & 4.79 \\
    \hline
    \rowcolor{lightgray}
    Guided Net  & 0.02662 & 31.67 & 0.933 & 0.987 & 3.17  & 3.90 \\
    \hline
    \rowcolor{lightgray}
    FeINFN      & 0.02412 & 32.74 & 0.948 & 0.983 & 2.85  & 3.94 \\
    \hline
    \rowcolor{lightgray}
    Fus Former  & 0.02616 & 31.89 & 0.943 & 0.985 & 3.14  & 3.97 \\
    \hline
    \rowcolor{lightgray}
    MIMO-SST    & \textbf{0.01632} & \textbf{35.85} & \textbf{0.965} & \textbf{0.992} & \textbf{2.12}  & \textbf{2.75} \\
    \hline
    Spectra Lift & \textit{0.01871} & \textit{34.79} & \textit{0.963} & \textit{0.991} & \textit{2.21}  & \textit{3.07} \\
    \hline
  \end{tabular}
\end{table}

\begin{table}[t]
  \centering
  \small
  \setlength\tabcolsep{2pt}
  \renewcommand{\arraystretch}{1.1}
  \captionsetup{font=small, justification=centering}
  \caption{Quality measures for the Pavia Center dataset without 1 band HR image: Mean value of 70 LR HSI-HR MSI configurations (\textbf{best in bold}, \textit{second best in italics}; supervised methods are shaded).}
  \label{tab:pavia_center_quality_without_1_band}
  \begin{tabular}{|l|c|c|c|c|c|c|}
    \hline
    \textbf{Method}
      & \textbf{RMSE $\downarrow$}
      & \textbf{PSNR $\uparrow$}
      & \textbf{SSIM $\uparrow$}
      & \textbf{UIQI $\uparrow$}
      & \textbf{ERGAS $\downarrow$}
      & \textbf{SAM $\downarrow$} \\
    \hline
    MIAE        & 0.03932 & 29.47 & 0.917 & 0.986 & 3.27  & 6.67 \\
    \hline
    C2FF        & 0.01770 & 35.36 & 0.978 & 0.996 & 1.73  & 3.43 \\
    \hline
    SDP         & 0.02081 & 33.72 & 0.971 & 0.995 & 2.02  & 4.31 \\
    \hline
    SSSR        & 0.05790 & 26.56 & 0.849 & 0.963 & 4.89  & 5.02 \\
    \hline
    \rowcolor{lightgray}
    Guided Net  & 0.02563 & 31.95 & 0.962 & 0.994 & 2.46  & 4.63 \\
    \hline
    \rowcolor{lightgray}
    FeINFN      & 0.02208 & 33.36 & 0.970 & 0.993 & 2.14  & 4.59 \\
    \hline
    \rowcolor{lightgray}
    Fus Former  & 0.01920 & 34.57 & 0.975 & 0.994 & 1.88  & 4.24 \\
    \hline
    \rowcolor{lightgray}
    MIMO-SST    & \textbf{0.01474} & \textbf{36.73} & \textbf{0.981} & \textit{0.996} & \textbf{1.56}  & \textbf{3.19} \\
    \hline
    Spectra Lift & \textit{0.01607} & \textit{35.93} & \textit{0.980} & \textbf{0.997} & \textit{1.61}  & \textit{3.30} \\
    \hline
  \end{tabular}
\end{table}

\begin{table}[t]
  \centering
  \small
  \setlength\tabcolsep{2pt}
  \renewcommand{\arraystretch}{1.1}
  \captionsetup{font=small, justification=centering}
  \caption{Quality measures for the Botswana dataset without 1 band HR image: Mean value of 70 LR HSI-HR MSI configurations (\textbf{best in bold}, \textit{second best in italics}; supervised methods are shaded).}
  \label{tab:botswana_quality_without_1_band}
  \begin{tabular}{|l|c|c|c|c|c|c|}
    \hline
    \textbf{Method}
      & \textbf{RMSE $\downarrow$}
      & \textbf{PSNR $\uparrow$}
      & \textbf{SSIM $\uparrow$}
      & \textbf{UIQI $\uparrow$}
      & \textbf{ERGAS $\downarrow$}
      & \textbf{SAM $\downarrow$} \\
    \hline
    MIAE        & 0.01725 & 35.53 & 0.962 & \textit{0.997} & 1.65  & 1.71 \\
    \hline
    C2FF        & 0.01428 & 37.19 & \textit{0.974} & \textbf{0.998} & \textit{1.34}  & \textit{1.46} \\
    \hline
    SDP         & 0.01742 & 35.32 & 0.955 & 0.995 & 2.21  & 1.88 \\
    \hline
    SSSR        & 0.01953 & 34.61 & 0.956 & 0.996 & 1.83  & 1.58 \\
    \hline
    \rowcolor{lightgray}
    Guided Net  & 0.01918 & 34.48 & 0.929 & 0.985 & 4.17  & 1.89 \\
    \hline
    \rowcolor{lightgray}
    FeINFN      & 0.02250 & 33.52 & 0.948 & 0.995 & 1.94  & 2.56 \\
    \hline
    \rowcolor{lightgray}
    Fus Former  & 0.02868 & 31.26 & 0.787 & 0.814 & 11.29 & 3.33 \\
    \hline
    \rowcolor{lightgray}
    MIMO-SST    & \textbf{0.01250} & \textbf{38.31} & \textbf{0.979} & \textbf{0.998} & \textbf{1.26}  & \textit{1.46} \\
    \hline
    Spectra Lift & \textit{0.01290} & \textit{37.80} & \textbf{0.979} & \textbf{0.998} & 1.47  & \textbf{1.31} \\
    \hline
  \end{tabular}
\end{table}
\clearpage

While SpectraLift achieves strong performance on most quality metrics (PSNR, SSIM, SAM, RMSE), it performs comparatively less well on ERGAS and UIQI in certain cases. This difference reflects the nature of these metrics: ERGAS and UIQI emphasize global luminance and contrast consistency, making them sensitive to small biases and variance shifts. In contrast, RMSE and SAM directly assess pixel-wise accuracy and spectral shape preservation, while SSIM captures spatial detail but focuses on local structural similarity. To maintain a lightweight, physics-grounded, and PSF-agnostic formulation, SpectraLift forgoes scene-specific PSF calibration and black-box spatial modeling—choices that can boost global metrics (ERGAS and UIQI) under synthetic conditions but often reduce robustness in real-world scenarios. Future work can extend this foundation to better address global consistency while preserving SpectraLift’s core strengths.

Beyond reconstruction accuracy, SpectraLift offers clear advantages in efficiency. Its training time is comparable to the fastest methods and substantially lower than that of more complex architectures, while its inference cost in FLOPs is close to the most lightweight approaches. This balance of high performance and low complexity underscores the practicality of SpectraLift for real-world hyperspectral super-resolution tasks. 

\begin{figure*}[!ht]
  \centering

  \begin{subfigure}{\textwidth}
    \includegraphics[width=\textwidth]{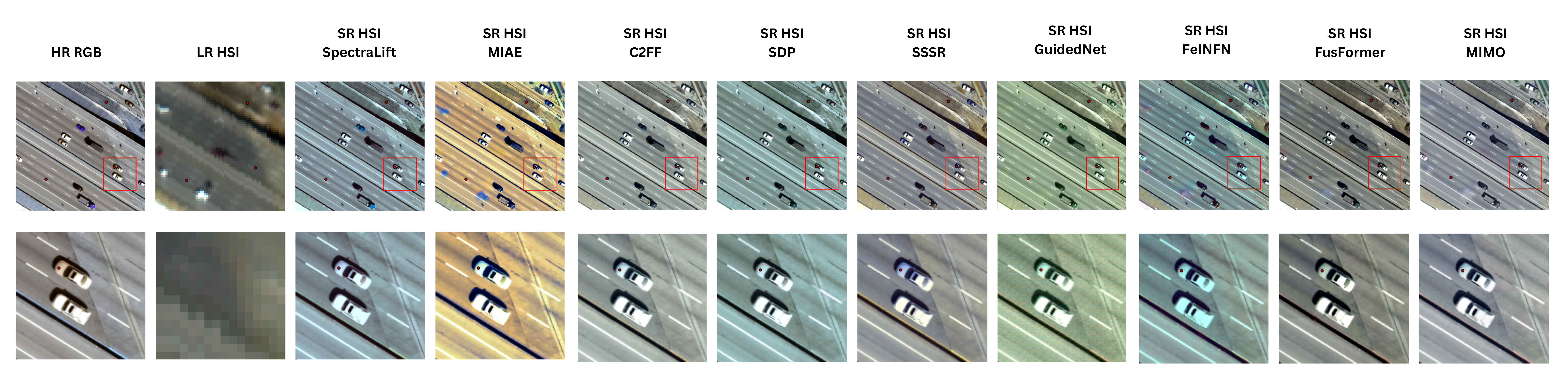}
    \caption{UH super-resolved images for test scene 1. Second row shows a zoomed-in crop of the region marked in red.}
    \label{fig:uh_bottomright}
  \end{subfigure}
  \hfill
  \begin{subfigure}{\textwidth}
    \includegraphics[width=\textwidth]{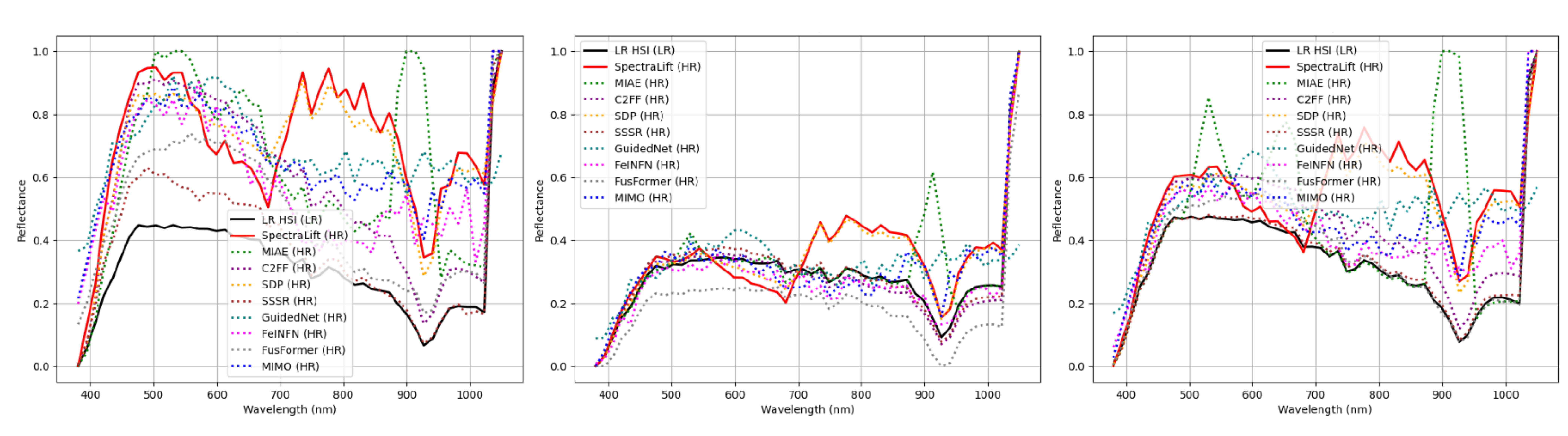}
    \caption{UH spectra for test scene 1. Left: Car, Center: Bare earth, Right: Highway.}
    \label{fig:uh_bottomright_spectra}
  \end{subfigure}

  \begin{subfigure}{\textwidth}
    \includegraphics[width=\textwidth]{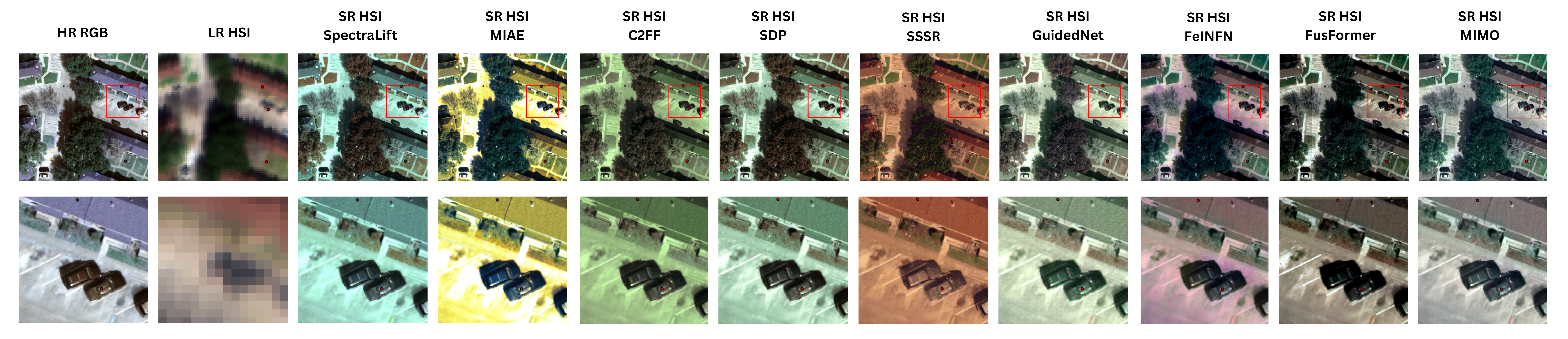}
    \caption{UH super-resolved images for test scene 2. Second row shows a zoomed-in crop of the region marked in red.}
    \label{fig:uh_topleft}
  \end{subfigure}
  \hfill
  \begin{subfigure}{\textwidth}
    \includegraphics[width=\textwidth]{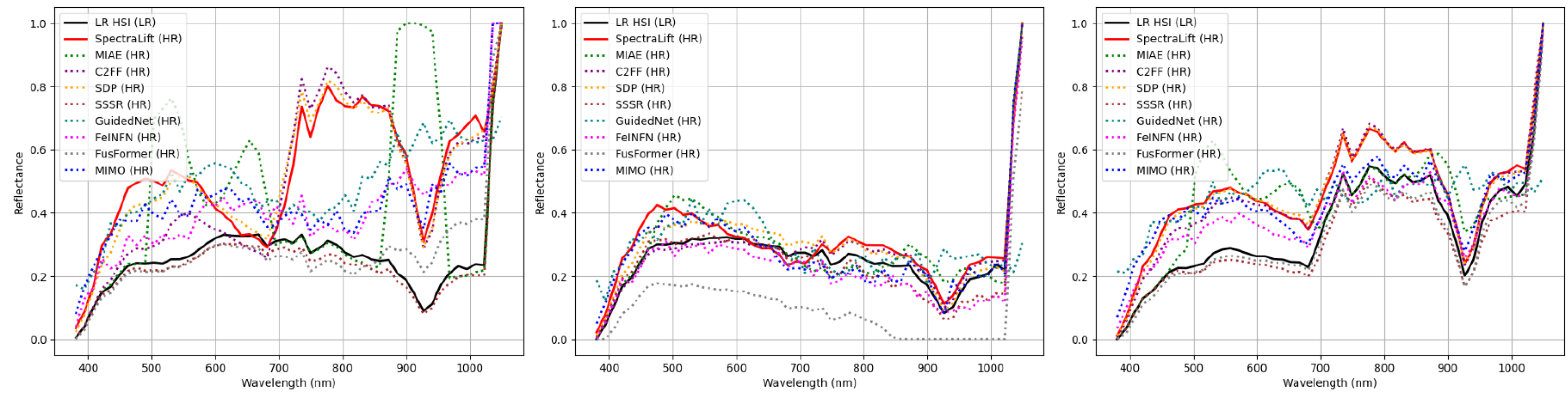}
    \caption{UH spectra for test scene 2. Left: Building roof, Center: Car, Right: Grass.}
    \label{fig:uh_topleft_spectra}
  \end{subfigure}

  \caption{University of Houston super-resolved results and corresponding spectra for two test scenes. (a, c) Super-resolved images with zoomed-in crops. (b, d) Spectral plots for selected regions.}
  \label{fig:uh_combined}
\end{figure*}

\subsection{Real-World Data}
We further evaluate SpectraLift on the University of Houston (UH) dataset from the 2018 IEEE GRSS Data Fusion Contest~\cite{UH_data}, which provides an LR-HSI ($4172\times 1202\times 50$) and an HR-RGB ($83440\times 24040\times 3$). We remove the last 2 bands from the LR-HSI due to noise, resulting in an LR-HSI of shape ($4172\times 1202\times 48$). We extract two distinct $64\times 64\times 48$ regions from the LR-HSI and their aligned $1280\times 1280\times 3$ HR-RGB patches, downsample the RGB to $512\times 512\times 3$ ($r=8$), and apply all unsupervised methods (including SpectraLift) to recover a $512\times 512\times 48$ HR-HSI. Testing is restricted to a small part of the UH dataset and the HR-RGB is downsampled from $1280\times 1280\times 3$ to $512\times 512\times 3$, in order to ensure that baseline methods could process the inputs in a single batch; if batch processing were used, the output of each batch would have to be stitched together and visible seam artifacts would appear (as can be seen for FusFormer in Fig.~\ref{fig:uh_bottomright} and ~\ref{fig:uh_topleft}, which could not process each region in a single batch). 

Since UH lacks HR-HSI ground truth and its PSF/SRF are unknown, we use the IKONOS RGB SRF and let each unsupervised method estimate the PSF. Supervised baselines were trained on Pavia Center HSI with inputs being generated by following Wald's protocol for $(r,c) = (8,3)$ via three overlapping $512\times 512\times 3$ HR-RGB patches and the corresponding LR-HSI patches, and then evaluated on the UH test pairs. For FusFormer, 53 non-overlapping HR-RGB patches of size $128\times 128\times 3$ and the corresponding LR-HSI patches had to be used to meet memory limits. 
Since no ground truth is available for this dataset and the true SRF, PSF is unknown, we avoid relying on QNR scores which are highly sensitive to these assumptions. We instead assess real-world performance via visual comparisons, generated by picking (Red = 659nm, Green = 561nm, Blue = 492nm) bands from the HR-HSI outputs of each method (Figs.~\ref{fig:uh_bottomright} and~\ref{fig:uh_topleft}), and spectral fidelity plots (Figs.~\ref{fig:uh_bottomright_spectra} and~\ref{fig:uh_topleft_spectra}).

Figure~\ref{fig:uh_bottomright_spectra} plots the spectral signatures at three manually selected pixels -- highway, bare earth, and car pixels -- marked in red in Figure \ref{fig:uh_bottomright}; Figure~\ref{fig:uh_topleft_spectra} does the same for grass, metal (car), and concrete (residential building) in Figure \ref{fig:uh_topleft}.  With a spatial downsampling factor of $r=8$, each LR-HSI pixel $\Y_{i,j,:}$ corresponds to an $8\times8$ region in the HR grid. To compare spectra, we pair the spectrum of the LR-HSI pixel at $(i,j)$ with that of the HR-HSI pixel at the top-right corner of the corresponding block, $(8i,,8j)$. For instance, in Figures~\ref{fig:uh_bottomright_spectra} and~\ref{fig:uh_topleft_spectra}, we compare the spectra of LR-HSI pixel $(8,8)$ and HR-HSI pixel $(64,64)$ to assess spectral fidelity.

\newpage Because the proprietary SRF of the RGB camera used to capture the HR-RGB images is unavailable, we adopt the IKONOS RGB SRF in our experiments. However, the IKONOS SRF is likely a poor approximation of the true sensor response, which remains unknown and cannot be reliably estimated without manufacturer-provided specifications. In practice, the true SRF is always known to the camera manufacturer and would be available for real-world missions, eliminating this limitation. Unfortunately, the manufacturer has not disclosed the SRF of this particular camera, preventing a more accurate approximation. This substitution introduces color shifts in the HR-HSI outputs, particularly affecting unsupervised methods that explicitly rely on the SRF. Supervised methods are also influenced, as their synthetic training data uses the same IKONOS SRF to generate LR-HSI pairs. If the true SRF were available—or if its characteristics (e.g., center, minimum, and maximum wavelengths of each band) were known, enabling a Gaussian or parametric approximation—this limitation could be alleviated.

These tints are most visible in GuidedNet, FeINFN, FusFormer, and MIMO-SST in Figure~\ref{fig:uh_bottomright}, and in MIAE, SDP, SSSR, and GuidedNet in Figure~\ref{fig:uh_topleft}. While SpectraLift is likewise affected— evidenced by red regions appearing grey (e.g., roofs, car, truck, and traffic cones)—it better preserves spatial structure and overall chromaticity than other baselines. Additional artifacts include checkerboard patterns in GuidedNet and patch-seam artifacts in FusFormer. Temporal misalignment between LR-HSI and HR-RGB, evident in Figure~\ref{fig:uh_bottomright} (e.g., moving cars), causes ghosting in joint-fusion methods (MIAE, FeINFN, FusFormer, MIMO-SST), whereas SpectraLift remains sharp and spatially coherent.

Across both UH scenes, SpectraLift’s per-pixel inversion recovers spectral shapes closely matching those of the LR-HSI, particularly for natural materials such as grass and bare earth (Figs.~\ref{fig:uh_bottomright_spectra}, \ref{fig:uh_topleft_spectra}). Although absolute reflectance values deviate—an expected outcome of the ill-posed spectral unmixing—the overall spectral structure remains well preserved, supporting downstream analysis. For certain man-made surfaces, such as highways and rooftops, other baselines achieve closer alignment with the LR-HSI spectra.

\subsection{Ablation}
\label{sec:ablation}

\begin{table}[t]
  \centering
  \small
  \setlength\tabcolsep{2pt}
  \renewcommand{\arraystretch}{1.1}
  \captionsetup{font=small, justification=centering}
  \caption{Ablation study of SpectraLift on the DC dataset using only Gaussian PSF \textbf{(best in bold)}.}
  \label{tab:ablation_table}
  \begin{tabular}{|l|c|c|c|c|c|c|}
    \hline
    \textbf{Model} 
      & \textbf{RMSE $\downarrow$} 
      & \textbf{PSNR $\uparrow$} 
      & \textbf{SSIM $\uparrow$} 
      & \textbf{SAM $\downarrow$}
      & \textbf{Params (M) $\downarrow$} 
      & \textbf{FLOPs (G) $\downarrow$} \\
    \hline
    Baseline                     & \textbf{0.0243} & \textbf{35.50} & 0.9652 & \textbf{3.3500} & 0.0336   & 26.28 \\
    \hline
    No skip connections          & 0.0247          & 35.29          & 0.9643 & 3.4155          & 0.0336   & 26.21 \\
    \hline
    No learning rate scheduler   & 0.0256          & 34.84          & 0.9622 & 3.4887          & 0.0336   & 26.28 \\
    \hline
    MSE loss                     & 0.0244          & 35.14          & 0.9603 & 3.4103          & 0.0336   & 26.28 \\
    \hline
    Cosine similarity loss       & 0.1549          & 17.15          & 0.7537 & 3.7929          & 0.0336   & 26.28 \\
    \hline
    ReLU                         & 0.0248          & 35.15          & 0.9649 & 3.3980          & 0.0336   & 26.28 \\
    \hline
    GeLU                         & 0.0250          & 35.04          & \textbf{0.9663} & 3.4001 & 0.0336   & 26.88 \\
    \hline
    8 hidden layers           & 0.0244          & 35.44          & 0.9650 & 3.3585          & 0.0420   & 32.79 \\
    \hline
    4 hidden layers           & 0.0244          & 35.47          & 0.9652 & 3.3690          & 0.0253   & 19.77 \\
    \hline
    2 hidden layers           & 0.0248          & 35.19          & 0.9661 & 3.3898          & 0.0170   & 13.25 \\
    \hline
    1 hidden layer           & 0.0267          & 34.53          & 0.9647 & 3.5704          & 0.0128   & 9.98 \\
    \hline
    32 hidden layer size               & 0.0246          & 35.16          & 0.9658 & 3.3561          & 0.0118   & 9.15 \\
    \hline
    128 hidden layer size              & 0.0276          & 34.68          & 0.9562 & 3.7911          & 0.1080   & 84.68 \\
    \hline
    Linear map                   & 0.0336          & 31.60          & 0.9623 & 4.2414          & \textbf{0.0012} & \textbf{0.90} \\
    \hline
  \end{tabular}
\end{table}

To quantify the impact of our architectural and training choices, we evaluate a suite of SpectraLift/SIN variants on the Washington DC Mall benchmark using only the Gaussian PSF; results are given in Table \ref{tab:ablation_table}. We perform the experiments using the same procedure as described in Section 3.2. All variants of SIN share the same hyperparameters and differ only in the component under test. Removing our selective skip/residual connections degrades reconstruction quality, confirming that skip connections stabilize the ill-posed spectral inversion by biasing the network when appropriate at almost no additional complexity cost. Disabling the learning-rate scheduler also harms convergence, while swapping the MAE loss for MSE causes a small drop in fidelity. By contrast, replacing MAE with a cosine-similarity loss collapses performance, underlining that we must penalize direct magnitude errors in the spectral domain. Changing activations -- Leaky ReLU to ReLU or GeLU -- has only a minor effect, showing robustness to the choice of nonlinearity.

SIN depth and width both matter: our chosen six-layer MLP achieves the best balance of accuracy and efficiency. An eight-layer variant offers no additional metric gains at substantially higher compute, whereas a four-layer or two-layer version remains competitive while having fewer parameters and FLOPs. Just using a MLP with a single hidden layer is also capable or achieving decent quality metrics at substantially lower computational cost. Reducing hidden-unit size further cuts complexity with only slight quality loss, while doubling hidden-unit size does not offer any benefits at much higher model complexity. A purely linear map (single dense layer without nonlinear activation) performs much worse, empirically validating that modeling hyperspectral signals on a near-linear manifold -- rather than as a strictly linear mapping -- is essential.

Overall, the baseline configuration is good for balancing the fidelity-efficiency trade-off, and users with tighter resource budgets can select the shallower or narrower variants to save compute with only marginal performance degradation.

\section{Conclusion}
\label{sec:conclusion}
We have introduced SpectraLift, a fully self-supervised HSI-SR framework that leverages the well-defined, readily available SRF (or its Gaussian approximations in cases where the exact SRF is not available) to directly invert spectral compression with a compact, per-pixel MLP. By casting fusion as an \(\ell_1\) spectral-inversion problem grounded in a clear physics model, SpectraLift avoids black-box approaches, and delivers interpretable mappings. Extensive evaluations across synthetic and real-world benchmarks show SpectraLift consistently matches or exceeds leading supervised methods on key quality metrics, while outperforming state-of-the-art unsupervised baselines. Its lightweight design and robustness to spatial blur and temporal misalignment further highlight its practical utility. SpectraLift establishes a foundation for lightweight, physics-guided spectral inversion and paves the way for future advances in efficient, transparent HSI-MSI fusion models for HSI-SR. Future research will focus on enhancing the interpretability of the inverse spectral mapping process, alongside the integration of context-aware mechanisms aimed at augmenting the single-band high-resolution input scenario, while preserving the foundational advantages of the SpectaLift framework. 

\section*{Acknowledgement}
We thank Dr. Mario Parente for many helpful discussions on practical assessments of HSI-SR methods.

\clearpage

\printbibliography

\end{document}